\documentclass[aps,twocolumn,showpacs]{revtex4}
\input epsf

\newcommand{\be}{\begin{equation}}
\newcommand{\ee}{\end{equation}}
\newcommand{\beq}{\begin{equation}}
\newcommand{\eeq}{\end{equation}}
\newcommand{\bea}{\begin{eqnarray}}
\newcommand{\eea}{\end{eqnarray}}
\newcommand{\beal}{\setcounter{letter}{1} \begin{eqnarray}}
\newcommand{\eeal}{\addtocounter{equation}{1} \end{eqnarray}}

\newcommand{\p}{\partial}

\begin{document}
%\draft

\title{\bf Anti-deSitter gravitational collapse}

\author{V. Husain${}^\flat$, G. Kunstatter$^\sharp$, B. Preston$^\sharp$ 
and M. Birukou${}^\sharp$}

\affiliation{$\flat$Dept. of Mathematics and Statistics\\
University of New Brunswick,
Fredericton, N.B. Canada E3B 1S5.\\
$\sharp$Dept. of Physics and Winnipeg Institute of
Theoretical Physics\\
University of Winnipeg,
Winnipeg, Manitoba Canada R3B 2E9.}

\date{\today}
%\maketitle

\begin{abstract}

We describe a formalism for studying spherically symmetric collapse of the  
massless scalar field 
in any spacetime dimension, and for any value of the 
cosmological constant $\Lambda$. The formalism is used for numerical 
simulations of 
gravitational collapse in four spacetime dimensions with 
negative $\Lambda$. We observe critical behaviour at the onset of black hole 
formation, and find that the critical exponent is independent of $\Lambda$.

\end{abstract}


\maketitle

It is now well established that gravitational collapse in spherical symmetry 
exhibits a phase transition-like critical behaviour, accompanied by 
self-similar behaviour of the matter field \cite{choptuik,gund,lehn}. 
The basic formula determined numerically for the black hole radius near the 
threshold of black hole formation is \cite{choptuik}.
\be 
    R_{BH} \sim (a-a_*)^\gamma,
\ee
where $a$ is an initial data parameter, and $a_*<a$ is its critical value.
The critical value $a=a_*$ gives a naked singularity, and illustrates a violation 
of the cosmic censorship conjecture. The exponent $\gamma$ is found to be 
``universal'' for a fixed matter type in the sense that it is  
independent of the functional form and parameter values of the initial matter 
profile. However it varies with the type of matter field. For perfect fluids
for example, $\gamma$ depends on the parameter(s) in the equation of state\cite{maison}.

A wide class of matter types yield two similar general features near 
criticality: the black hole radius scaling law (without a mass gap), 
and discrete or continuous self-similarity of the matter fields. (These are 
called Type II transitions. There are also transitions with a mass gap, 
refered to as Type I, where the field is either periodic or static, rather 
than self-similar.)

One of the questions that has not been studied to date is that of the 
dependence of the critical exponent on  the cosmological 
constant $\Lambda$. We study this question in four spacetime dimensions. 
(The case of three dimensions is rather special in 
that  there are no
static black hole solutions unless the cosmological constant is 
negative;
the constant may be scaled to unity without loss of generality so there is
no need to study such dependence in this case \cite{pc,ho}). Our main result is 
that $\gamma$ is independent of $\Lambda$. 

A useful by-product of our analysis is confirmation of the utility of
a new formalism for studying the collapse problem, in which the $d-$dimensional, 
spherically symmetric  Einstein-scalar equations are rewritten as an effective 
$2-$dimensional 
dilaton gravity theory\cite{dil_grav}. All spacetime and $\Lambda$ 
information in this approach is stored in the resulting dilaton potential. This feature 
permits writing a ``universal''
numerical code in which spacetime dimension and 
$\Lambda$ appear as input parameters.  In fact the formalism is general enough 
to  permit simulations of mass and potential terms for the scalar field, 
with minimal changes to the code. 

The reduced equations are written in double null coordinates, and a numerical 
method first used by Goldwirth and Piran\cite{GP}, and refined by Garfinkle\cite{DG} 
is implemented on the resulting equations. The code is in principle capable of handling 
any spacetime dimension and cosmological constant value, although there are practical 
constraints. We describe below the main features of the formalism and numerical 
method. Further details appear in Ref. \cite{nd}.  

Einstein gravity with cosmological constant and minimally 
coupled scalar field in $d$ spacetime dimensions is given by the 
action
\bea
S^{(d)}&=&{1\over 16\pi G^{(d)}}\int d^dx\sqrt{-g^{(d)}}
   \left[R(g^{(d)}) - \Lambda\right] \nonumber \\
   && - \int d^dx\ \sqrt{-g^{(d)}}|\p\chi|^2.
\label{Einstein}
\eea
Spherical symmetry is imposed by writing the metric $g_{\mu\nu}$ as
\be
ds^2_{(d)} = \bar{g}_{\alpha\beta} dx^\alpha dx^{\beta} + r^2(x^\alpha)
d\Omega_{(d-2)},
\ee
where $d\Omega_{(d-2)}$ is the metric on $S^{d-2}$ and $\alpha,\beta =
1,2$.

A useful form for the reduced action with this form of the metric is
obtained by defining $l= (G^{(d)})^{n/2}$ and 
\bea 
\phi &:=& {n\over 8(n-1)}\left({r\over l}\right)^n, \\
g_{\alpha\beta} &:=& \phi^{2(n-1)/n}\ \bar{g}_{\alpha\beta}, 
\label{defs}
\eea
where $n\equiv d-2$. Note that the $\phi$ is proportional to
the area of an $n$-sphere at fixed radius $r$. With these definitions 
the reduced action becomes 
\bea
S &=& {1\over 2G}\int d^2x \sqrt{-g}\left[\phi R(g) +
V^{(n)}(\phi,\Lambda)\right]
\nonumber \\
&& -\int d^2x\sqrt{-g}\ H^{(n)}(\phi)\ |\partial \chi|^2
\label{dilatonaction}
\eea
where $G\equiv 2\pi n/(n-1)$, the scalar field has been rescaled
($\chi \to \chi/\sqrt{G^{(d)}})$ in order to make it dimensionless, 
and the overall factor of the $n-$sphere volume has been dropped 
from the action. 
In addition we have defined 
\bea
H^{(n)}(\phi) &\equiv& {8(n-1)\over n} \phi, \\
V^{(n)}(\phi,\Lambda) &\equiv& {1\over n} \left[ {8(n-1)\over n} \right]^{1\over n} 
\phi^{1\over n}
\nonumber \\
&&\times  
\left[
{n^2\over 8} \left({8(n-1)\over n}\right)^{n-2\over n}
\phi^{-{2\over n}} -l^2 \Lambda\right] 
\label{poten}
\eea

Now with the metric parametrized as
\beq
ds^2 = - 2 l g(u,v)\phi'(u,v) du dv
\label{double null}
\eeq
the field equations are
\bea
& &\dot{\phi}' = - {l\over 2} V^{(n)}(\phi) g \phi'
\label{double null equations a}\\
& &{g'\phi'\over g H^{(n)}(\phi)} = 2G (\chi')^2 \label{cons}\\
& &(H^{(n)}(\phi)\chi')^{\cdot} + (H^{(n)}(\phi) \dot{\chi})' = 0,
\label{double null equations c}
\eea
where prime and dot denote the $v$ and $u$ derivatives respectively.

The evolution equations may be put in a form more useful for numerical
solution by defining the variable
\be h = \chi + {2\phi\chi' \over \phi'},
\label{def h}
\ee
which replaces the scalar field $\chi$ by $h$. The evolution equations become
\bea
\dot{\phi} &=& -\tilde{g}/2 \label{phidot}\\
\dot{h} &=& {1\over 2\phi} (h - \chi)\left( g\phi V^{(n)} - {3\over 2}\tilde{g}\right),
\label{hdot}
\eea
where
\bea
g &=& \exp\left[4\pi\int^v_udv{\phi'\over\phi}(h-\chi)^2\right],  \\
\tilde{g} &=& l\int _u^v (g \phi'V^{(n)}) dv,
\label{gbar}
\eea
\be
\chi={1\over
2\sqrt{\phi}}\int^v_udv\left[{h\phi'\over\sqrt{\phi}}\right].
\label{chi integral}
\ee
This is the final form of the equations used for numerical evolution. 
The spacetime parameters $n$ and $\Lambda$ appear only
in the dilaton 
potential $V^{(n)}$ (\ref{poten}).

% 
%\section{Numerical Method} 
%

The numerical scheme uses a $v$ discretization to obtain a
set of coupled ODEs:
\be
h(u,v) \rightarrow h_i(u),\ \ \ \ \ \ \ \phi(u,v)\rightarrow \phi_i(u).
\ee
where $i = 0,\cdots, N$ specifies the $v$ grid. Initial data for these
two
functions
is prescribed on a constant $v$ slice, from which  
$g(u,v),\tilde{g}(u,v)$ are
constructed. Evolution in $u$ (`time')
is performed using the 4th. order Runge-Kutta method. The general 
scheme is similar to that used in \cite{GP}, with some 
refinements used in \cite{DG}. This procedure was also used for the 
$3-$dimensional collapse calculations in \cite{ho}.

The initial scalar field configuration $\chi(\phi,u=0)$ is most
conveniently specified as a function of $\phi$ rather than $r$. (Recall 
that $\phi\propto r^n$.)  This together with the initial arrangement of 
the radial points $\phi(v,u=0)$ fixes all other functions. We used the initial
specification $\phi(0,v) = v$.

Most of the computations are for the initial scalar field  of 
the Gaussian form
\be
\chi_G(u=0,\phi) = a \phi\ {\rm exp}\left[-\left(\ {\phi-\phi_0\over
\sigma}\right)^2\right],
\ee 
with  attention restricted to variations of the amplitude $a$. However we 
also used the ``cosh'' initial data 
\be 
\chi_G(u=0,\phi) = a {\rm cosh} [b(\phi-\phi_0)]^{-2},
\ee
to study convergence of our code and to test universality with 
non-zero $\Lambda$.
 
The initial values of the other functions are determined in terms of the 
above by computing the integrals for $g_n$ and $\tilde{g}_n$ using Simpson's 
rule.

The boundary conditions at fixed $u$ are
\be
\phi_k=0, \ \ \ \tilde{g}_k=0,\ \ \ \ g_k=1.
\ee
where $k$ is the index corresponding to the position of the origin
$\phi=0$. (In the algorithm used, all grid points $ 0\le i \le k-1$
correspond to ingoing rays that have reached the origin and
are dropped from the grid). These conditions are equivalent
to $r(u,u)=0$, $g|_{r=0} = g(u,u) =1$, and guarantee regularity of
the metric at $r=0$.  For our initial data, $\phi_k$ and
hence $h_k$ are initially zero, and therefore remain zero at the origin
because of Eqn. (\ref{hdot}).

At each $u$ step, the function 
\be
ah\equiv {g}^{\alpha\beta}\p_\alpha \phi \p_\beta \phi =
-{\dot{\phi}\over lg},
\label{aheqn}
\ee
is observed. Its vanishing signals the formation of an apparent horizon.
For each run of the code with fixed amplitude $a$, this function is
scanned from larger to smaller radial values after each  Runge-Kutta
iteration, and evolution is terminated if the value of this function
reaches $10^{-4}$. (The code cannot continue much beyond 
this value.) The corresponding radial coordinate value is recorded
as $R_{ah}$. In the subcritical case, it is expected that all the radial
grid points reach zero without detection of an apparent horizon. This 
is the signal of pulse reflection.
The results $(a, R_{ah})$ are collated 
by seeking a relationship of the 
form
\beq
R_{ah}\propto (a-a_*)^{\gamma}.
\label{sl}
\eeq

The code was tested for grid sizes ranging from 2000 to 20000 points, and
with the $u$ and $v$ step sizes ranging from $10^{-2}$ to $10^{-6}$, for
the two types of initial data used, as well as the vacuum case  $\chi=0$.
These tests established that the code converges. 
 
For most of the runs we used  $\phi_0=1$ and $\sigma =0.3$ for the
Gaussian initial data, and $b=5.0$, $\phi_0=0.5$ for the cosh data. 
We varied $\phi_0$ in the Gaussian data from 0.1 to 20 (which also varies 
pulse width in the $r$-coordinate), to see if there was $\gamma$ dependence. 
This is important to study because $\Lambda$ sets a 
scale in the problem.

Our results for the two types of data (for $n=2$) appear in Tables 1 and 2, which 
list the 
computed $\gamma$ values for the corresponding (negative) values of 
$\Lambda$. The trend is clear: $\gamma$ {\it does not depend on} $\Lambda$. 
The first value provides a crucial test of our formalism and code since it 
reproduces the known $\Lambda = 0$ value of the exponent \cite{choptuik}. 
\medskip

\begin{center} 
\begin{tabular}{lcr}
\hline
\hline
$\ \ \ \Lambda$  && $\gamma\ \ \ \ \ $ \  \\
\hline
-0.001 && $\ \ \ \ \ \ \ \ \ \ \  $0.370-0.375 \\
-5 &&  0.37-0.38 \\
-10 && 0.37-0.39 \\
-20 && 0.36-0.38 \\
\hline
\hline  
\end{tabular}
\end{center} 
\centerline{Table 1. Computed ranges of $\gamma$ for Gaussian data ($n=2$).}
\bigskip

\begin{center} 
\begin{tabular}{lcr}
\hline
\hline
$\ \ \ \Lambda$  && $\gamma\ \ \ \ $ \  \\
\hline
-5 && $\ \ \ \ \ \ \ \ \ \ \ \ \ \ \ \ \ $0.37-0.38 \\
-20 && 0.36-0.38 \\
-50 && 0.37-0.40\\
\hline
\hline  
\end{tabular}
\end{center} 
\centerline{Table 2. Computed ranges of $\gamma$ for cosh data ($n=2$).}
\bigskip

The ranges of $\gamma$ in Tables 1 and 2 are arrived at by 
assessing our uncertainty in the determination of the critical 
amplitudes $a*$: the $a*$ values are determined to lie 
within certain domains, and the end points of these domains are 
used to determine the  range of $\gamma$ values. 
For example, for $\Lambda = -20$ and cosh data, we find a* lies in the range 
$6.00344 \times 10^{-3}$ -- $6.00360\times 10^{-3}$, which gives $\gamma$ in the 
range indicated in Table 2. Our determination of the $a*$ window gets 
coarser with increasing $\Lambda$, which leads to the larger error bars 
on the $\gamma$ values,
as indicated in the tables.

Figure 1 is the ln-ln graph  of the apparent horizon 
radius vs initial data amplitude, for $\Lambda = -5$. This is representative 
of the type of results from which the critical exponent ranges in 
Tables 1 and 2 are deduced.

%FIG 1
\begin{figure}[t]
\begin{center}
\epsfxsize=80mm
\epsffile{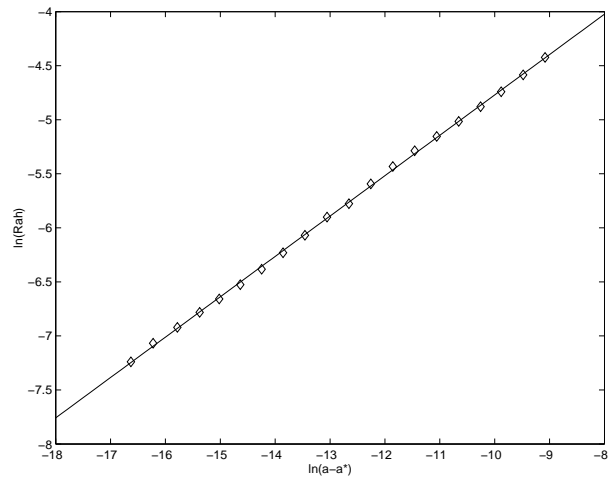}
\caption{\baselineskip = 1.0em Logarithmic plot of apparent horizon
radius $R_{ah}$ versus initial scalar field amplitude $(a-a_*)$ for 
$\Lambda = -5$. The line is the least squares fit 
to the points giving $\gamma = 0.3738$}
\end{center}
\end{figure}

%FIG 2
\begin{figure}[t]
\begin{center}
\epsfxsize=80mm
\epsffile{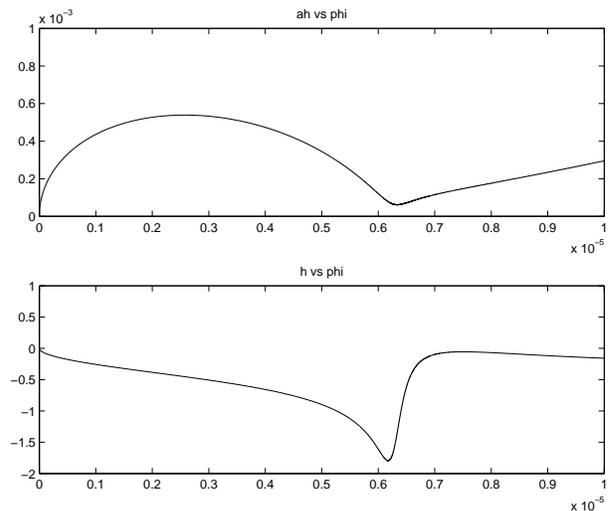}
\caption{\baselineskip = 1.0em  Typical graphs of the apparent horizon 
function (\ref{aheqn}) and the scalar field $h$ near black hole 
formation.}
\end{center}
\end{figure}  

Figure 2 shows plots of the apparent horizon function 
(\ref{aheqn}) and the scalar field $h$. This is typical of the 
type of data used to extract horizon radius information at the 
supercritical onset of black hole formation. These graphs  show 
impending black hole formation at $\phi \sim 6.3\times 10^{-6}$ 
($r \sim 5\times 10^{-3})$ for a supercritical amplitude with 
$\Lambda=-5$. 

The dip in the apparent horizon function (Fig. 2) oscillates 
toward and away from the origin as it approaches the $\phi-$axis, and 
appears to be the cause of the small oscillations about the least 
squares fit line in Figure 1. Although these latter oscillations have 
been noted in earlier work, it appears that they are the manifestation 
of the dip oscillations in the apparent horizon function (\ref{aheqn}) 
near criticality. This in turn is connected with the discrete self-similarity 
of the scalar field, which not surprisingly, is also manifested in the 
apparent horizon function. Figure 1 contains two complete oscillations, 
and may be used to obtain a rough estimate of the echoing period:  
it gives $\Delta \sim 3.5$. This value is close to the $\Lambda=0$ value 
$\Delta = 3.44$ computed in earlier work\cite{choptuik,gund,lehn}. 
Thus, both the numbers $\gamma$ and $\Delta$ associated with critical 
behaviour appear to be independent of $\Lambda$. 
 
It is known, and verified again here, that $\gamma$ 
is 
independent of initial data profiles and parameter values.  We also 
find that for fixed $\Lambda$, critical exponents are independent of 
characteristic pulse width and starting location in relation to the 
$\Lambda$ scale.   

It is worth noting that in our equations $\Lambda$ appears only in the 
potential term (\ref{poten}), where for sufficiently small $\phi$, the $\Lambda$ 
term is insignificant. This provides some analytical support of our results, 
since it suggests that the entire near critical part of the evolution, with  
typical $\phi$ range $0-10^{-5}$ (Fig. 2), is $\Lambda$ independent. 
In four dimensions, $n=d-2=2$, the first term in brackets in (\ref{poten}) is 
of order $10^5$, which dominates the $\Lambda$ term for all cases we consider. 
An extrapolation of this observation, beyond numerical reach, suggests that 
$\gamma$ is $\Lambda$ independent because sufficiently close to criticality, the 
$\phi$ term  dominates the $\Lambda$ term for any $\Lambda$ (in an $\epsilon -  
\delta$ sense).  

It is interesting to contrast this with the minimally coupled scalar field of 
mass $\mu$, which also has a scale. Here two types of critical 
behaviour are observed at the threshold of black hole formation, depending on 
initial pulse width $\sigma$ in comparison with $\mu^{-1}$ \cite{brady}: for 
$\sigma >> \mu^{-1}$ there is a mass gap, whereas for $\sigma << \mu^{-1}$ there 
is no mass gap and the exponent $\gamma \sim 0.378$ is computed\cite{brady}. 

There are some differences here worth emphasizing. The first is specific to our 
results, and the second is general: (i) We find no evidence of a mass gap for a 
range of values of pulse width and initial position in relation to the 
scale $1/\sqrt{-\Lambda}$. With no mass gap, the
critical exponent is measured 
in the limit of small horizon values. In  our simulations, the
typical  
horizon size range at the onset of black hole formation is $10^{-2}-10^{-4}$. 
By contrast, the smallest $\Lambda$ scale the code can handle is 
$1/\sqrt{50}\sim 0.14$. Under these circumstances, it is not surprising that 
$\Lambda$  has no effect. (ii) Unlike $\mu$, $\Lambda$ appears in the stress-energy 
tensor independent of the scalar field. Therefore it is not unreasonable that 
$\Lambda$ and $\mu$ give qualitatively different results.
 
For postive $\Lambda$ we are not able to obtain accurate results for 
$\gamma$ because of two competing effects in our procedure: since the grid 
is evolving,  we observe an outward  deSitter expansion of the grid, which 
confines the interesting features of the ingoing collapse to an ever shrinking 
region near the origin. Perhaps this case can better studied by the method of 
replacing lost grid points used in \cite{DG}, and using the subcritical 
method of computing $\gamma$ using the curvature scalar,  introduced in \cite{DG2}. 
Nevertheless, on the intuitive grounds mentioned,
we expect that the sign of 
$\Lambda$ will not change our results. 
 
It is known that critical exponents may be computed via linear perturbation 
analysis of the critical solution \cite{maison,kha}. This has so far been done
only for $\Lambda=0$. For $\Lambda\ne 0$, it is reasonable to expect that both 
the critical solution and its perturbation equation depend on $\Lambda$, which 
after all is a parameter in the equations of motion. It would be useful to see 
how $\Lambda$ drops out of this type of calculation, yielding a $\Lambda$ 
independent exponent. 

In summary, our numerical simulations of massless scalar field collapse 
in spherical symmetry in four dimensions, show that the critical exponent
associated with the collapse is independent of (negative) $\Lambda$ values. 
This result extends the scope of universality to include 
the cosmological constant, and suggests that including $\Lambda$ with 
other matter types, such as the perfect fluid or the Yang-Mills field, 
will also not change the critical exponent.

%\section{Acknowledgements}

This work was supported in part by the Natural Sciences and Engineering
Research Council of Canada.  We thank Rob Myers and Eric Poisson 
for a discussion. 

  \par
\end{document}